\documentclass[11pt]{article}
\usepackage[utf8]{inputenc}
\usepackage[T1]{fontenc}
\usepackage{url}
\usepackage{indentfirst}
\usepackage{graphicx}
\graphicspath{{./plots/}}
\usepackage{subfig}
\usepackage{float}
\usepackage{amsmath}
\usepackage{amsfonts}
\usepackage{amssymb}
\usepackage{longtable}
\usepackage{fancyvrb}
\usepackage{paralist}
\usepackage{booktabs}
\usepackage{array}
\usepackage{graphicx}
\usepackage{wrapfig}
\usepackage[colorlinks=false,urlcolor=black,linkcolor=black]{hyperref}

\pdfoutput=1
\usepackage{hyperref}
\pdfoutput=1
\begin{document}
\title{Electrohydrodynamically induced mixing \\ in immiscible multilayer flows}
\author{Radu Cimpeanu, Demetrios T. Papageorgiou \\
\\\vspace{6pt} Department of Mathematics, \\ Imperial College London, SW7 2AZ London, United Kingdom}
\maketitle
\begin{abstract}

In the present study we investigate electrostatic stabilization mechanisms acting on stratified fluids. 
Electric fields have been shown to control and even suppress the Rayleigh-Taylor instability when a heavy fluid lies above lighter fluid. 
From a different perspective, similar techniques can also be used to generate interfacial dynamics in otherwise stable systems.
We aim to identify active control protocols in confined geometries that induce time dependent flows in small scale devices without having moving parts. 
This effect has numerous applications, ranging from mixing phenomena to electric lithography. 
Two-dimensional computations are carried out and several such protocols are described.
We present computational fluid dynamics videos with different underlying mixing strategies, which show promising results.
\end{abstract}
\section{Introduction}

The field of microfluidics has been one of the most active areas in fluid dynamics
for the past few decades. With applications as diverse as microchip design
and medical/pharmaceutical devices, recent advances in theoretical, numerical and experimental 
settings have had a powerful impact in the research world.

A key process in such systems is represented by mixing of agents, which becomes increasingly
challenging as lengthscales become smaller and reach micrometer-sized geometries. Very low
Reynolds numbers generate numerous difficulties in the control of such devices and
several passive or active mechanisms to manipulate the fluid flow in an accurate way have been explored to date.
In the following paragraphs, we focus especially on electrohydrodynamically controlled models, which have
proved successful in reaching remarkable results with limited resource consumption.

An introduction of the model
of the effects of an electric field on a flow of immiscible fluids
in a channel is introduced by Ozen et al.\cite{ozen1} Linear stability theory, as well
as a variety of theoretical parameter studies centered around a Reynolds number of $1$, which is typical in microfluidic context, are presented. 
The impact of the electric field, as well as other quantities in the problem such as initial 
position of the interface within the channel or viscosity ratio are discussed. An experimental setup
from the same authors \cite{ozen2} is constructed in order to identify features of drop formation
in a channel as a result of the influence of electric fields of various strengths.
The results reported are based on a channel of dimensions $70$ mm (in $x$) $\times$ $0.25$ mm (in $y$) $\times$ $1.5$ mm (in $z$)
with a background Poiseuille flow at a flux which generates a $Re=\mathcal{O}(10^{-2})$. Glycerine
and corn oil have been used as the two immiscible fluids in the system. Key findings indicate
how drop size decreases as the prescribed voltage is increased. Another extensive theoretical
and numerical study of instabilities in a channel flow that can be used for mixing applications
is shown in Ozen et al. \cite{ozen3}. Scenarios with both Couette and Poiseuille flow are considered
for leaky, as well as perfect dielectrics. Computations are carried out for a large set of
values of the Reynolds numbers($0-10^4$), as interesting discussions can be based on parameter regimes
around known critical Reynolds numbers for the classical flows. An important result is given by the fact
that in the case of perfect dielectrics, the electric field normal to the interface always has
destabilizing effects, which can then be exploited in the context of microfluidic mixing.

Lee et al. \cite{lee1} have recently conducted a highly acclaimed review study of the most successful mixing devices in geometries
pertaining to microfluidic flow. Key parameters in the vast majority of contemporary water-based systems
are of $Re=\mathcal{O}(10^{-1})$, with reference lengthscales of the order of $100\ \mu$m. These magnitudes provide an estimate which
allows us to design a theoretical framework, as well as a computational study with applicability to devices presently used. The authors
also indicate the experimental work of El Moctar, Aubry and Batton \cite{moctar1} as representative for systems based on electrohydrodynamic forces.
El Moctar et al. use a T-type mixer with fluids of similar properties (in this case corn oil, however dyed in a different color and with different electric properties 
in each inflow channel)
of sizes $30$ mm (in $x$) $\times$ $0.25$ mm (in $y$) $\times$ $0.25$ mm (in $z$) and subject to an electric field corresponding to approximately $10^5$ V/m.
The setup corresponds to a Re $<0.02$ and both continuous and alternating currents have been used with results
drastically improving over scenarios with no electric field. T-shaped mixers are in general
one of the most popular choices for mixing devices (\cite{moctar1},\cite{goullet1},\cite{tsouris1},\cite{glasgow1},\cite{johnson1},\cite{lu1})
due to the their richness of experimental and modelling possibilities and hence versatility for parameter studies resulting in rapid advances for this application.
The use of time pulsing \cite{goullet1} has shown to be particularly successful in this context, leading to high degrees of mixing
on shorter timescales. Reynolds numbers are again in the order of $10^{-1}-10^{1}$ ($0.3$ and $2.55$ for the mentioned publication)
and reference lengthscales are $\mu$m-sized. Electric fields strengths for such geometries are of the order $10^5-10^6$ V/m, which
is very common for the relevant microdevices. Another example can be found in the experiment of Tsouris et al.\cite{tsouris1},
where flows characterized by Re$=0.2,0.4$ and $0.9$ are subjected to electric fields of $0-2\cdot 10^6$ V/m and show 
highly improved mixing as the electric field strength increases.

In the present work we aim to describe a mixing mechanism that requires no hydrodynamic forcing or a certain
imposed velocity field. Instead we rely on control protocols targeted towards the electric field only. The interfacial
dynamics achieved in response to electric excitation is then proved to be effective in terms of reaching
high degrees of mixing efficiency. 
Due to simplicity and small resource consumption,
the protocols we describe become an attractive alternative to classical choices in microgeometries. Note that
similar mixing policies can then be applied to further enhance the performance of existing devices in a very broad context.

\section{Mathematical Description}

The mathematical framework on which we construct our study is similar
to the investigation of Cimpeanu, Papageorgiou and Petropoulos\cite{cimpeanu1}, focused on
the electrohydrodynamic stabilization of the Rayleigh-Taylor instability in an infinite vertical channel.

\begin{figure}[!ht]
\centering
\includegraphics[width=12cm]{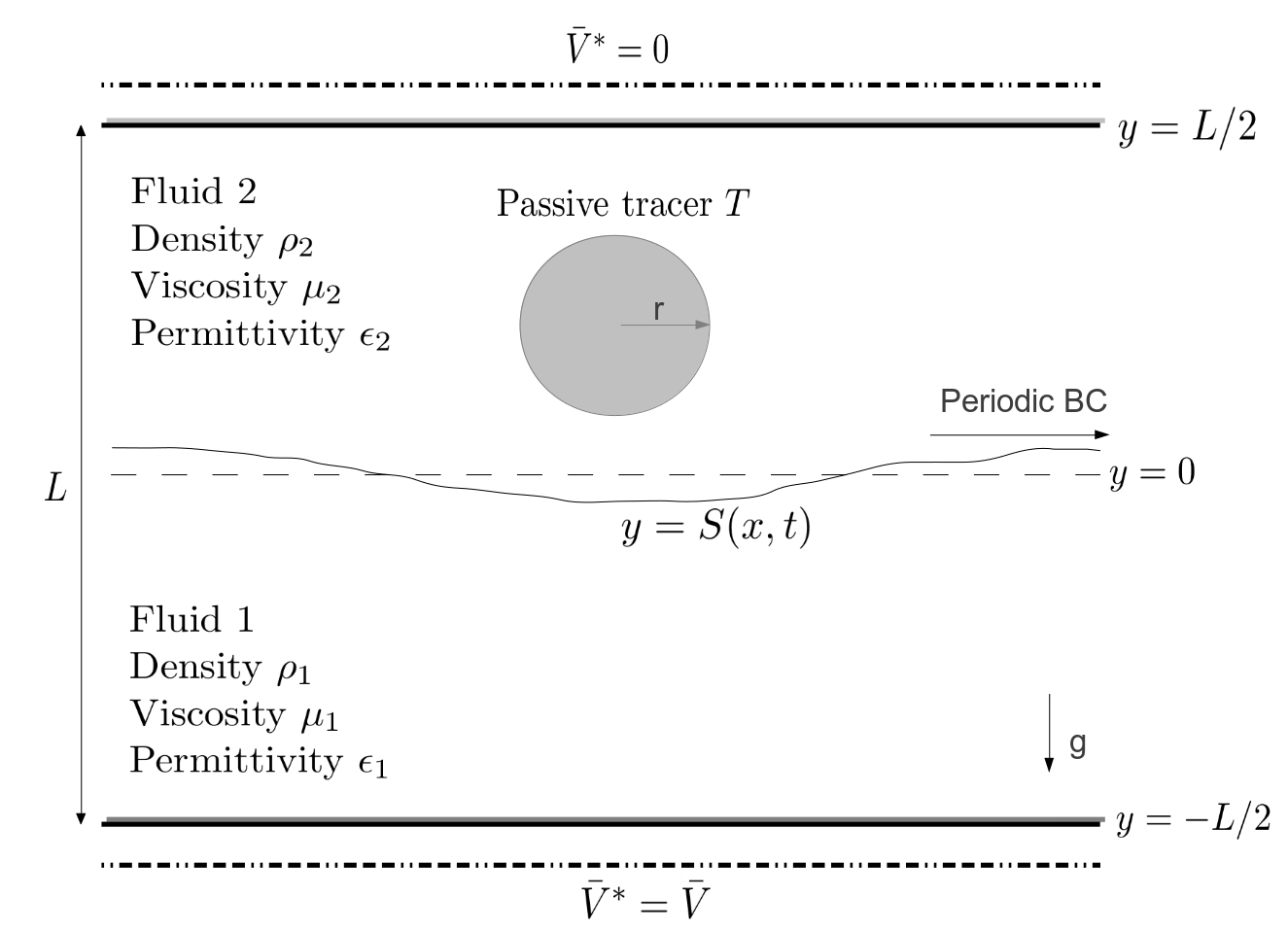}
\caption{\label{fig:fp_Schematic} Sketch of domain}
\end{figure}

In the present work we consider two incompressible, immiscible, viscous fluids in a two-dimensional setting as shown in Fig.~\ref{fig:fp_Schematic}.
The flow is bounded by horizontal parallel walls that are separated by a distance $L$, and are unconfined in the lateral
direction as shown in the figure (periodic boundary conditions are considered).
Using a Cartesian coordinate system, the interface between the two fluids is denoted by $y=S(x,t)$, and fluids 1 and 2 occupy
the regions $y<S(x,t)$ and $y>S(x,t)$, respectively (in what follows subscripts 1,2 will refer to fluids 1 and 2).
The horizontal walls at $y = \pm L/2$ are no-slip, no-penetration boundaries and are also electrodes that can support a voltage potential difference.
The fluids are perfect dielectrics with given permittivities $\epsilon_{1,2}$, viscosities $\mu_{1,2}$ and densities $\rho_{1,2}$, and
corresponding velocity vectors are $\textbf{u}_{1,2}=(u_{1,2},v_{1,2})$. 
We denote the constant surface tension coefficient at the interface
by $\sigma$. 

An electric field is imposed by grounding the electrode at $y=L/2$ and imposing a constant voltage $\bar{V}^*$ at $y=-L/2$.
The voltage potentials $V_{1,2}$ in regions 1,2 satisfy Laplace's equation (this follows from the electrostatic approximation:
Maxwell's equations reduce to $\nabla\times\textbf{E}_{1,2}=0$, $\nabla\cdot(\epsilon_{1,2}\textbf{E}_{1,2})=0$, hence 
$\textbf{E}_{1,2}=-\nabla V_{1,2}$ from the former condition with Laplace equations following from the second condition away from the interface):
\begin{equation}
\left(\frac{\partial^2 }{\partial x^2}+\frac{\partial^2 }{\partial y^2}\right)V_{1,2} = 0.\label{eq:Laplace}
\end{equation}
The dimensional momentum and continuity equations are
\begin{eqnarray}\label{eq:dimensionalNSE}
 \rho_1 (\textbf{u}_{1t}+(\textbf{u}_1 \cdot \nabla)\textbf{u}_1) &=& - \nabla p_1 + \mu_1 \Delta \textbf{u}_1 - \rho_1 g \textbf{j}, \\ 
 \rho_2 (\textbf{u}_{2t}+(\textbf{u}_2 \cdot \nabla)\textbf{u}_2) &=& - \nabla p_2 + \mu_2 \Delta \textbf{u}_2 - \rho_2 g \textbf{j},\\
  \nabla \cdot{{\textbf{u}}_{1,2}} &=& 0.
 \end{eqnarray}
We introduce the density, viscosity and permittivity ratio parameters 
\begin{eqnarray}
r=\rho_1/\rho_2,\ m=\mu_2/\mu_1,\ \epsilon = \epsilon_2/\epsilon_1,
\end{eqnarray}
and non-dimensionalize the equations and boundary conditions using fluid $1$ as reference.
Lengths are scaled by $L$, velocities by a reference value $U$ and pressures by $\rho_1 U^2$.
We list the following dimensionless parameters
\begin{equation}
 \tilde{g}=\dfrac{gL}{U^2},\ \tilde{\mu}=\dfrac{\mu_1}{\rho_1 U L}\ W_e=\dfrac{\sigma}{\rho_1 g L^2},
\end{equation}
representing an inverse square Froude number $\tilde{g}$, an inverse Reynolds number $\tilde{\mu}$
and an inverse Weber number denoted by $W_e$. Note that since we consider $U \sim \sqrt{gL}$,
dimensionless number $\tilde{g} \sim 1$ for all cases. The effect of gravity can be artificially
increased or decreased by modifying the value of this number, however usually gravity plays a negligible role within
devices of very small physical lengthscales.

Furthermore, we scale voltage potentials by $V^*$ so that the dimensionless
electric parameter measuring the size of Maxwell stresses in the interfacial stress balance equation becomes unity in fluid 1 variables.
Inspection of the stress tensor shows that this choice necessitates
\begin{eqnarray}
\rho_1 U^2=\frac{\epsilon_1 (V^*)^2}{L^2}\Rightarrow V^*=UL\sqrt{\rho_1/\epsilon_1}.\label{eq:Vstar}
\end{eqnarray}

With these scalings the Navier-Stokes equations for each fluid become
\begin{eqnarray}\label{eq:dimensionlessNSE}
 \tilde{\textbf{u}}_{1t}+(\tilde{\textbf{u}}_1 \cdot \nabla)\tilde{\textbf{u}}_1 &=& - \nabla \tilde{p}_1 + \tilde{\mu} \Delta \tilde{\textbf{u}}_1 - \tilde{g} \textbf{j}, \\ 
 \tilde{\textbf{u}}_{2t}+(\tilde{\textbf{u}}_2 \cdot \nabla)\tilde{\textbf{u}}_2 &=& - r \nabla \tilde{p}_2 + m\tilde{\mu} r \Delta \tilde{\textbf{u}}_2 - \tilde{g} \textbf{j},
 \end{eqnarray}
where $\textbf{j}$ is the unit vector in vertical direction and the decoration tilde is used
to refer to dimensionless quantities. The continuity equation in each fluid is
\begin{equation}
 \nabla \cdot{\tilde{\textbf{u}}_{1,2}} = 0.
\end{equation}

From the previously described set of equations, following a classical linearization procedure and normal mode analysis, 
we identify the most unstable wavenumbers within a certain setup. We concentrate on stably stratified formats, where
the vertical electric field can be used to generate and enhance instabilities.
Exploiting this, we construct initial perturbations
that allow for the rapid formation of high amplitudes of the disturbance and ultimately lead to efficient mixing.
This effect is achieved by imposing on-off protocols in the electric field, which simply means oscillating
between a uniform vertical electric field to destabilize the flow, followed by an interruption of the voltage feed. 
The repeated use of such a control leads to rich dynamics of the passive tracer.

In the on-off scenario, the voltage is controlled via the boundary condition on $\bar{V}^{*}$ at $y=-1/2$ (after non-dimensionalization).
This can either be a positive prescribed constant $\bar{V}$ for the "on"-mode, whereas the "off"-mode is described by $\bar{V}^{*} = 0$ at $y=-1/2$.
We notice (see the first segment of accompanying simulation video) that the dynamics generated by the vertical motion of the interface is sufficient to achieve 
high degrees of mixing. More interestingly however, it is possible to generate horizontal motion as well (see right side of first segment of the attached video), since
the mechanism tries to select the most unstable mode at the expense of breaking symmetry in the current interfacial shape. The 
following electric field manipulation is geared towards controlling this particular effect.

The alternative to the on-off protocol is the imposition of a relay-type structure, where the time-dependent voltage is now described by
\begin{equation}
 \bar{V}^{*} = \bar{V} + a ( (f \cdot (\textrm{atan}(x+s \cdot t) - x_0)) - (f \cdot (\textrm{atan}(x+s \cdot t) - x_1))).
\end{equation}
Notation $a$ is used for the amplitude of the voltage fluctuation, which is normalized by $\pi$, $\bar{V}$ is the imposed background voltage,
$s$ is a term that gives the velocity of the leftward or rightward moving perturbation, while $f$ is the factor that 
controls the arctan-smoothing. A high value of $f$ results in a very steep slope of the disturbance, whereas a small value of $f$
leads to well-behaved transition from $\bar{V}$ to $\bar{V}+a$ over a larger area. This perturbation in the voltage is then contained between regions $x_0$
and $x_1$, which need to have appropriately chosen values within our scaling. Multiple such perturbations are constructed to mimic the structure of the
most unstable wavenumber as picked up by linear stability and allow the generation of time-dependent flows within our confined geometry.
This is essentially a form of microfluidic pumping, which will be investigated in more detail in the near future.

All simulations have been performed using the GERRIS \cite{popinet1} package, which employs the volume-of-fluid method
to discretize the multi-fluid system. Several other features such as spatial adaptivity and parallelization, as well as numerical techniques
specialized for solving the Navier-Stokes equations, are available in order to optimize the numerical treatment of the problem.

\section{Key Parameters}

The first segment of the attached simulation video (roughly $45$ seconds) is composed of three simulations stacked horizontally, 
each representing a separate on-off protocol scenario. All parameters
related to the fluids themselves are kept the same, the only difference is the non-dimensional time at which the electric field is turned on or off.
The imposed voltage is constant in all computational experiments and is set to $\bar{V} = 6.0$.

The domain has non-dimensional size $1\times 1$, while the relevant fluid parameters are
\begin{itemize}
 \item Density ratio $r = \rho_1/\rho_2 = 6/1$;
 \item Viscosity ratio $ m =\mu_2/\mu_1 = 1/10$;
 \item Dimensionless viscosity $\tilde{\mu} = 0.025$;
 \item Permittivity ratio $\epsilon = \epsilon_2 / \epsilon_1 = 2/1$;
 \item Surface tension $\tilde{\sigma} = 0.1$;
 \item Passive tracer radius $r = 0.1$;
 \item Enhanced dimensionless gravity $\tilde{g}=10.0$.
\end{itemize}

We allow the simulations to run over approximately six dimensionless time units and the spatial adaptivity is set to allow
for a maximum of $2^8=256$ cells in the case of all variables in the problem, except for the interface and the horizontal velocity,
which can carry a maximum of $2^9=512$ cells, thus resulting in a minimum $h = 1/512 \approx 0.002$. 

The imposed electric field in each of the simulations (from left to right) is as follows:
\begin{itemize}
 \item Left: on between $t=0.0 - 5.0$, off between $t=5.0-6.0$;
 \item Center: on between $t=0.0 - 1.0$, $t=2.0 - 3.0$ and $t=4.0 - 5.0$, off between $t=1.0 - 2.0$, $t=3.0 - 4.0$ and $t=5.0 - 6.0$;
 \item Right: on between $t=0.0 - 2.0$ and $t=4.0 - 6.0$, off between $t=2.0 - 4.0$.
\end{itemize}

The animation shows the concentration field $T$ varying from 0 (blue) to 1 (red), with a circular initial condition. The aim
is to reach a homogeneous structure inside the concentration field, as a result of the mixing procedure. In white we show 
the active fluid interfacial shape, with an initial perturbation of amplitude $0.025$ and a wavenumber of $k=6\pi$. As the electric field is turned on,
the perturbation grows and generates motion affecting the passive tracer. The switching off of the electric field then allows the stabilization
to a flat interface. Repeating this procedure within a certain range of appropriate parameters becomes an effective mixing strategy.

The second segment of the attached video contains three examples of relay type constructions. As in the previous case, the geometry and fluid properties are kept the same,
the differences lie in the amplitude of the voltage perturbation and the imposed (horizontal) velocity of this anomaly. The fluids are characterized by
the same set of properties as before, however the electric fields are now given by a base voltage of $\bar{V}=6.0$ with either
\begin{itemize}
 \item Left: voltage perturbation amplitude $a = 1.0$, velocity $s=1.0$;
 \item Center: voltage perturbation amplitude $a = 1.5$, velocity $s=1.0$;
 \item Right: voltage perturbation amplitude $a = 1.5$, velocity $s=0.5$.
\end{itemize}
An additional feature of the GERRIS package, droplet removal, has been used to limit numerical artifacts in the solution. Furthermore,
in black we plot equipotential lines, which allow for the clean visualization of the imposed voltage as a function of time. As non-dimensional
time $t$ reaches $1$ unit, the relay is switched on and this type of microfluidic pumping generates a flux that initiates the horizontal motion of the interface.
The mixing of the passive tracer becomes highly efficient in this case and can be further enhanced by combining this strategy with on-off protocols as described before.

All simulations are modelled to contain fictitious fluids, with properties that are representative in the context of our study. Fully
realistic values, representing two actual fluids, as could be reproduced under experimental conditions, will be used at a later stage.
The Reynolds number in the computational experiments is of order $\mathcal{O}(10^1)$ and requires further reduction as we enter the microfluidic range.

\section{Future Work}

Identifying optimal mixing protocols in a general framework is the main focus of the project at the current stage.
Once satisfactory results are obtained, we will direct our attention to micro-scale devices, where
existing strategies will be improved to adapt to low Reynolds number flows. Realistic fluids, frequently 
used in experimental contexts, will be preferred to the current model. Further extensions to other
geometries (such as a full T-mixer) and three-dimensional generalizations are also within reach.

\end{document}